\documentclass[preprint,prb,showkeys,showpacs,tightenlines]{revtex4}
\usepackage{graphicx}
\begin{document}
\preprint{Bicocca-FT-05-12  March 2005}

\title
 {The $\phi_3^4$  lattice field theory viewed from  \\
   the high-temperature side\\}

\author{P. Butera\cite{pb} and M. Comi}
\affiliation
{Istituto Nazionale di Fisica Nucleare and\\
Dipartimento di Fisica, Universit\`a di Milano-Bicocca\\
 3 Piazza della Scienza, 20126 Milano, Italy}
\date{\today}
\begin{abstract}

We analyze  high-temperature series expansions of the two-point and 
four-point correlation-functions in the 
 three-dimensional euclidean lattice scalar field 
theory with quartic self-coupling, which have been recently extended  
through twenty-fifth order for the simple-cubic and body-centered-cubic 
lattices.     
We conclude that the length of the present series is sufficient for a fairly 
accurate description of the critical behavior of the model and 
confirm the validity of universality, scaling and hyperscaling.
In the case of the body-centered-cubic lattice, we determine the value 
of the quartic self-coupling for which the leading 
corrections to scaling approximately  vanish and correspondingly 
the universal critical parameters can be determined with high accuracy.
In particular, for the susceptibility and the correlation-length exponents
we find $\gamma=1.2373(2)  $ and $\nu=0.6301(2) $.
For the four-point renormalized coupling we find $g=23.56(3) $.
In the case of the simple-cubic lattice our results are consistent with 
 earlier estimates.
\end{abstract}


\maketitle

\section{Introduction}

 High-temperature (HT) expansions for  scalar spin models
 with bilinear nearest-neighbor interaction and general single-spin 
measure, on 2- and 3-dimensional
bipartite lattices, have been  
extended\cite{dyca,bin,nr90,rosac,bcesse25,pisa25} 
 in the last few decades from
 order $K^{10}$ to $K^{25}$.  Here $K=\frac {J} {kT}$ 
denotes the usual HT expansion variable,  with 
$k$ the Boltzmann constant, $T$ the temperature
 and $J$ some energy scale characterizing 
 the next-neighbor interaction.
The main observables expanded, 
include the susceptibility
$ \chi_2$, the second moment of the correlation-function $\mu_2$, the
 zero-momentum four-point function $ \chi_4$.
 The linked-cluster method used in these calculations expresses 
 the series coefficients in a closed form as polynomials in the 
(normalized) moments of the single-spin measure thus making the analysis
 of a broad class of models possible.
This fact entails two related benefits. First, it enables one to 
 explore the extent of the Ising universality class to which these
models should generally belong. Second,
one can take advantage of the non-universality 
of the amplitudes of the corrections to scaling 
 in the critical behavior to obtain very accurate estimates
 of universal critical parameters, such as exponents and scaling functions,
 by focusing the numerical analysis on those 
particular spin models, within the given 
universality class, which  show vanishing (or quite small)
 leading corrections to scaling(LCS).
This prescription, which can  be effective only with 
rather long series, was 
suggested in Refs.\cite{zinn81,fisher} and extensively 
tested by 21-term HT series on the body-centered-cubic lattice, 
for various single-spin measures.

 In the initial studies, however, the available series were simply too short.
In Ref.\cite{bin} an  analysis of the critical properties
of the lattice $\phi^4$ euclidean field theory 
 in two-, three- and four-dimensional space, using tenth-order HT series,
   suggested a complicated and puzzling critical
behavior in the 3d case, 
 which showed failures of universality and hyperscaling and cast
 doubt on the validity of key mathematical assumptions\cite{schra} 
( see however Ref.\cite{bagn}) in the 
application of the renormalization group to the critical phenomena.
 This analysis  was   historically  important 
in raising  questions which could be answered only by numerical 
investigation of specific models and was a strong incentive to 
further extend the HT series and to repeat and improve
simulation studies.
The anomalous features observed in Ref.\cite{bin} 
were not confirmed by the successive studies\cite{nr90,fisher,succ} 
and  were finally ascribed to the shortness 
 of the series  analyzed. Therefore  over a decade ago the problem
 was considered as settled. 

 We have recently extended, through order $K^{25}$, the HT expansions
of  the moments of the correlation-function
and, through  order $K^{23}$, the expansion of the zero-momentum 
four-point correlation-function
 for a general scalar  model defined on  three-dimensional
 bipartite lattices (such as the simple-cubic (sc) and the 
body-centered-cubic (bcc) lattices).
These expansions  have already proved useful for  accurately displaying 
  the properties of critical scaling and universality  of the 
spin-$S$ Ising model\cite{bcesse25,bcessetab25}
 and have also produced high-precision estimates of critical exponents
 and  universal combinations of critical amplitudes, based on the 
prescriptions of Refs.\cite{zinn81,fisher}. 
 Also following 
the lead of the same Authors\cite{zinn81,fisher}, 
analogous expansions, independently 
obtained\cite{pisa25} for the sc lattice case, 
through order 25 for $\chi_2$ and $\mu_2$ and 
 through order 21 for $\chi_4$, 
  were used
 to study the  lattice $\phi^4_3$ model in a neighborhood of a 
particular value of the self-coupling 
for which the LCS are negligible.
 It remains to show that the  present HT expansions 
 are also adequate  for an extensive and 
accurate study of the scalar field model for two different lattices and
  on the whole range of the self-coupling.
From our study we  conclude 
 that the present HT series  are sufficiently long   that, 
 by properly resumming them,   
 our analysis can reach an accuracy comparable 
or higher  than the most extensive  MonteCarlo simulations. 
Moreover,  we show that also 
in the case of the bcc lattice, we can  approximately determine 
  a value of the quartic 
self-coupling such that the LCS  vanish,  and thereby  
we can obtain\cite{zinn81,fisher}
 very accurate estimates 
of the universal critical parameters in complete agreement with 
 the other cited recent computations\cite{bcesse25,pisa25}.

In the following paragraphs we shall first recall  the definitions 
 of the quantities whose HT expansions are  studied.
We shall then very briefly comment on the results of the series analysis.
We shall map out the phase diagram
 of the model and exhibit  its universality properties 
 along the critical line,
 such as the  independence of the critical exponents $\gamma$ and $\nu$
 on the self-coupling  of the field and on the lattice structure.
Then, we shall give evidence that the critical 
renormalized coupling constant is  universal and  non-zero, 
  thus verifying the validity of hyperscaling.
 Finally we shall report our best estimates of the critical 
exponents $\gamma$ and $\nu$ and of the renormalized four-point coupling 
from the analysis of the  model with minimal LCS on the bcc lattice.
 The figures  summarizing our results are among the main motivations of this 
 study, since some of them qualitatively differ from 
   the analogous ones appearing in Ref.\cite{bin},
  which, to the best of our knowledge,  have never been
 revised in the literature.

\section{ The  model}

 The lattice $\phi_3^4$ model is defined by the Hamiltonian
\begin{equation}
K {\it H}\{\phi\}=-K \sum_{<i,j>} \phi_i \phi_j
 +\sum_i(\phi_i^2 +g_0(\phi_i^2-1)^2)
\label{Hamilt} \end{equation}
Here  $i$ and $j$ are   integer-component (multi)indices denoting
 the  sites of a three-dimensional lattice and the
 first sum extends to all nearest-neighbor sites.

This Hamiltonian is obtained from the lattice discretization of the euclidean
 action of the continuum real field $\psi(x)$ in $2+1$ dimensions.
\begin{equation}
 S= \int  [ \frac {1} {2} (\partial_{\mu} \psi)^2 + \frac {1} {2}m^2 \psi^2
 + \frac {\lambda} {4!}\psi^4] dx
\label{Act} \end{equation}
after setting $\psi = \sqrt{2K} \phi$,  $ m^2= \frac{1-2g_0} {K} -6$ and 
$\lambda = \frac{6 g_0} {K^2}$

We will analyze the HT expansion of the moments of the two-point connected
 correlation-function $<\phi_i \phi_j>_c $ and of the 
zero-momentum connected four-point correlation-function
$ \chi_4(K,g_0) = \sum_{i,j,k} < \phi_0 \phi_i \phi_j \phi_k >_c$.

The susceptibility is defined by the zero-order moment of $<\phi_0 \phi_i>_c$
\begin{equation}
 \chi_2(K,g_0) = \sum_i <\phi_0 \phi_i>_c .
\label{Sus} \end{equation}
 Its behavior, as $K$ tends from below to the critical value $K_c(g_0)$, 
 is expected to be 
\begin{equation}
 \chi_2(K,g_0) = A(g_0) \tau(g_0)^{-\gamma(g_0)}
[ 1 + a(g_0)\tau^{\theta(g_0)} +...] 
\label{Suscrit} \end{equation}
where $A(g_0)$ is the critical
 amplitude of the susceptibility, 
$\tau (g_0) = 1-K/K_c(g_0)$  is called the reduced (inverse) temperature, 
 $\gamma(g_0)$ is the critical exponent of the susceptibility,
$a(g_0)$ is the amplitude of the LCS  
and $\theta(g_0)$ is the exponent of the LCS.

The square of the correlation-length  is expressed in terms of
 the ratio of the second 
moment of the correlation-function
 $ \mu_2(K,g_0)= \sum_i i^2 <\phi_0 \phi_i>_c$ 
and the susceptibility as 
\begin{equation}
\xi^2(K,g_0) =
  \frac  {\mu_2(K,g_0)} {6 \chi_2(K,g_0)}
\label{Corl} \end{equation}
and is expected to show the critical behavior
\begin{equation}
 \xi^2(K,g_0) = B(g_0) \tau(g_0)^{-2\nu(g_0)}
[ 1 + b(g_0)\tau^{\theta(g_0)} +...] 
\label{corcrit} \end{equation}
 where $\nu(g_0)$ is the critical exponent of the  correlation-length.

The critical renormalized coupling constant $g(g_0)$ 
is expressed in terms of 
$ \chi_4(K,g_0)$,   $ \chi_2(K,g_0)$ and $\xi^2(K,g_0)$  as the 
value of the quantity
\begin{equation}
 g(K,g_0)= \frac {-v \chi_4(K,g_0)} {\xi^{3/2}(K,g_0)  \chi_2^2(K,g_0)}
\label{gren} \end{equation}
when $K$ tends from below to  $K_c(g_0)$.
Here $v$ denotes the volume per lattice site. We have $v=1$ for the sc
lattice and $v=4/3\sqrt 3$ for the bcc lattice.
The expected critical behavior of 	$g(K,g_0)$ is 
\begin{equation}
 g(K,g_0)= g(g_0) \tau(g_0)^{\gamma(g_0)+3\nu(g_0)-2\Delta_4(g_0)}
[ 1+ c(g_0)\tau^{\theta(g_0)} +...]
\label{grencrit} \end{equation}
 where $\Delta_4(g_0)$ is the gap exponent.
If $\gamma(g_0)+3\nu(g_0)-2\Delta_4(g_0) = 0$, we say that 
hyperscaling holds and the critical renormalized coupling
 $g(g_0)$ is the finite positive \cite{glja} critical limit of
 $g(K,g_0)$.
Of course, the exponents $\gamma(g_0)$, $\nu(g_0)$, $\Delta_4(g_0)$,
 $\theta(g_0)$ 
 and the critical renormalized coupling  $g(g_0)$, as 
determined from the analysis of the HT series, 
will appear to be independent of $g_0$,  if  universality 
is valid along the critical line.

\section{ Numerical results}

\subsection {The phase diagrams}
\begin{figure}
\includegraphics[width=3.5in]{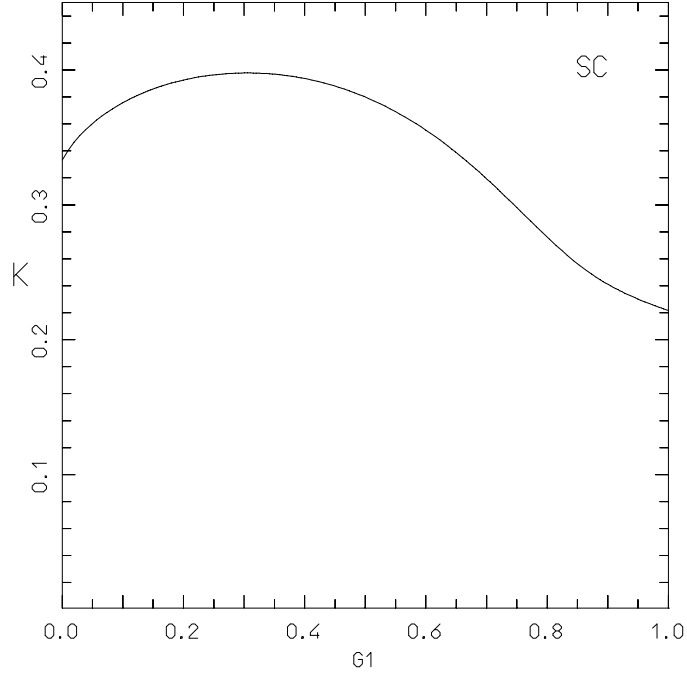}
\caption{\label{fig_1_phi43} Phase diagram of  $\phi_3^4$ on the sc lattice.
 The critical line $ K=K(g_0)$ is plotted vs. $G1= \frac {g_0} {g_0+1}$.}
\end{figure}

\begin{figure}
\includegraphics[width=3.5in]{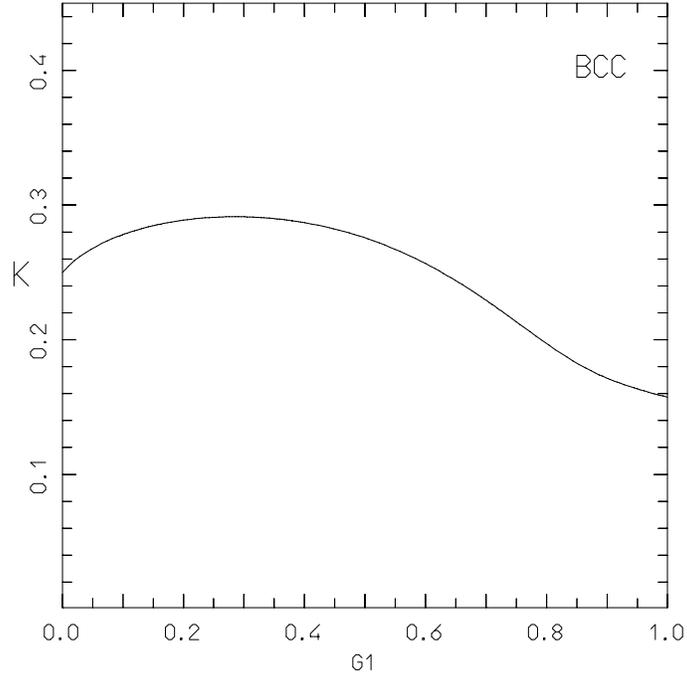}
\caption{\label{fig_2_phi43} Phase diagram of  $\phi_3^4$ on the bcc lattice.
 The critical line $K=K(g_0)$ is plotted vs. $G1= \frac {g_0} {g_0+1}$.}
\end{figure}

In Figs. \ref{fig_1_phi43}  and \ref{fig_2_phi43}  
we have mapped out the phase diagrams of the $\phi_3^4$
 model on the sc and the bcc lattices, respectively.
There are  two phases separated by a line $K=K_c(g_0)$
 of second-order critical points.
In the disordered phase, below the critical line,  the reflection symmetry
$\phi \rightarrow -\phi$ is unbroken.
For $g_0= 0 $ the model reduces to the Gaussian model and  we have 
$K_c(0)=2/q=1/3$ on the
sc lattice ($K_c(0)=2/q=1/4$ on the bcc lattice), where $q$ is the 
lattice coordination number. 
 For $g_0 \rightarrow \infty $ the model reduces to the usual 
 spin-$1/2$ Ising model and we have $K_c(\infty)=0.221655(2)$
on the sc lattice ($K_c(\infty)=0.1573725(10)$ on the bcc lattice)
 as indicated in our general study\cite{bcesse25} of the spin-$S$ Ising model.
In Table 1 we have reported  numerical estimates of 
$K_c(g_0)$   for a few 
values of $0<g_0<\infty$ in the sc and the bcc lattice cases.
 These estimates are obtained by the following  
simple procedure. 
Using  the susceptibility  
expansion, for each value of $g_0$,
  we  form   the sequence $\{K^{(n)}_c(g_0)\}$ 
of Zinn-Justin modified-ratio  approximants (MRA)\cite{zinn81,guttse}
 of the critical point. We observe that these sequences 
 approach smoothly their expected asymptotic  behavior\cite{bcesse25} 
\begin{equation} 
K^{(n)}_c(g_0) =K_c(g_0)\big [1 - 
C(\gamma,\theta)a(g_0)/n^{1+\theta(g_0)} + o(1/n^{1+\theta(g_0)})\big ] 
\label{fitbeta} \end{equation}
where $a(g_0)$ and $\theta(g_0)$ are, respectively, the amplitude
 and the exponent of the LCS
 appearing in eq.(\ref{Suscrit}) and 
  $C(\gamma,\theta)$ is a positive
constant\cite{bcesse25}  depending  on the values 
 of $\gamma(g_0)$ and $\theta(g_0)$.
We expect that the exponent $\theta(g_0)$ 
  be universal, namely independent of $g_0$ 
 and indeed, to a good approximation,
 the asymptotic behavior  of the MRA  sequences
 is consistent with 
the value $\theta= 0.517(4)$   suggested by a recent simultaneous
 study\cite {blode} of a set of models in the Ising universality class.
 Since, for all values of $g_0$, 
the contributions of the higher-order terms in eq.(\ref{fitbeta}) appear
 to be  small, particularly so in the bcc lattice case, we shall 
assume that the MRA sequences $\{K^{(n)}_c(g_0)\}$ 
can be  extrapolated  to infinite length $n$ of the series 
simply by fitting the highest-order (alternate) approximants to
 the first two terms of the asymptotic expansion (\ref{fitbeta}).
 The values of $K_c(g_0)$ determined by this prescription are
 quite stable, both under small 
variations of $\theta$  and  of the fitting procedure, such as including
 in the fit higher-order terms of the expansion (\ref{fitbeta}).
Our conclusion is that  at least five decimal figures of the result
  are reliable.
 These  estimates are then refined 
by comparing the results of the extrapolations with the determination 
 of $K_c(g_0)$ from first-, second- and third-order differential 
approximants(DA) \cite{guttse,gutasy} and finally error bars
 are  inferred which are likely to be only  
 upper bounds.
 As expected,  these uncertainties, which reflect
 the differences between the estimates by extrapolated 
MRA sequences and by DAs, depend strongly on the size of the 
corrections to scaling and therefore on the
 value of $g_0$, but of course  they 
  are generally invisible on the scale of  figures \ref{fig_1_phi43} 
and \ref{fig_2_phi43}.

\subsection {The critical exponents}

\begin{figure}
\includegraphics[width=3.1in]{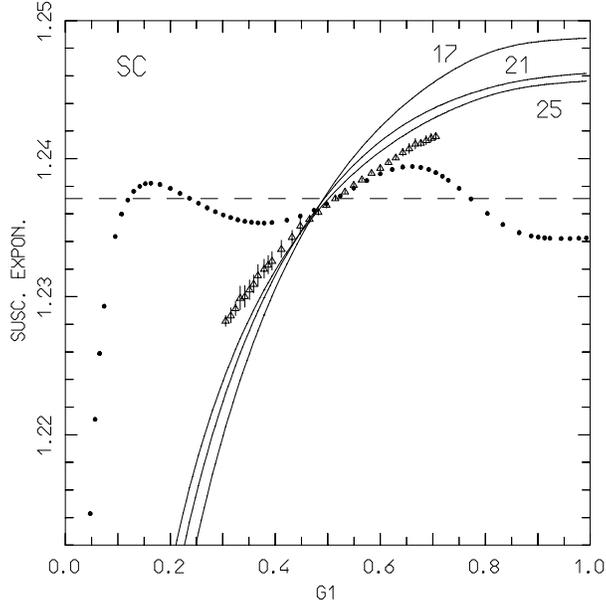}
\caption{\label{fig_3_phi43} The susceptibility exponent $\gamma(g_0)$ 
 as a function of $G1 = \frac {g_0} {g_0+1}$   for the sc lattice.
 The continuous curves represent the  estimates obtained by the  
 MRAs  of orders $17$, $21$ and $25$ respectively. 
The dashed curve indicates the central value of our 
estimate\cite{bcesse25} of $\gamma$ for the Ising universality class.
The dotted curve shows the results of an extrapolation of the
MRA sequences by a fit
 to the asymptotic expansion (\ref{fitgamma})
supplemented by  a single  higher-order correction $O(1/n^4)$.
The triangles show  estimates of $\gamma(g_0)$ by 
 simple unbiased first-order DAs.}
\end{figure}

\begin{figure}
\includegraphics[width=3.1in]{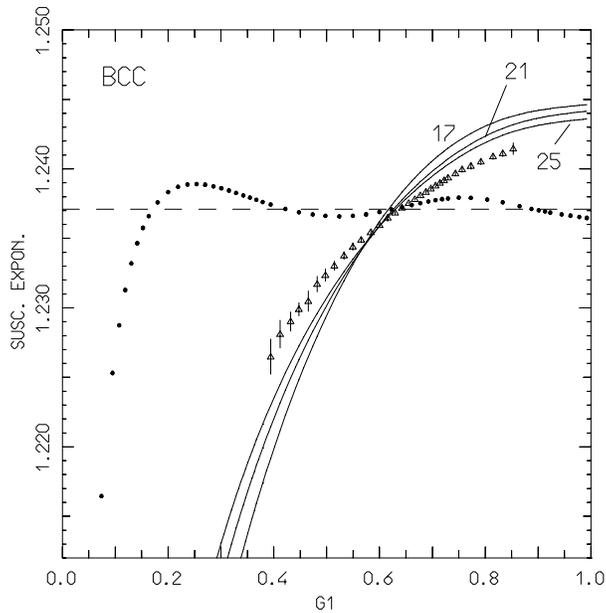}
\caption{\label{fig_4_phi43} 
The same as in Fig.\ref{fig_3_phi43}, but for the bcc
lattice.}
\end{figure}

In Figs. \ref{fig_3_phi43}  and \ref{fig_4_phi43}, 
referring to the sc and the bcc lattices respectively,
we have plotted our estimates of the exponent of the susceptibility 
$\gamma(g_0)$ vs.  $G1=
\frac {g_0} {g_0+1}$. Also in this case, 
 using the susceptibility HT expansion,
 for each value of $g_0$, we can form the sequences 
$\{\gamma^{(n)}(g_0)\}$ of MRAs of the  exponent
$\gamma(g_0)$ and
compare them  to 
their expected asymptotic expansion\cite{bcesse25}
\begin{equation} 
\gamma^{(n)}(g_0) =\gamma(g_0)-D(\gamma,\theta) a(g_0)/n^{\theta(g_0)}+ 
 o(1/n^{\theta(g_0)}). 
\label{fitgamma} 
\end{equation}
Here $ D(\gamma,\theta)$ is a positive constant\cite{bcesse25} 
depending on $\gamma(g_0)$ and $\theta(g_0)$
 and $a(g_0)$ is  the amplitude of the LCS in eq.(\ref{Suscrit}).
We can observe that also these MRA  sequences are smooth, 
but a simple minded extrapolation based on a fit of the 
 sequences $\{\gamma^{(n)}(g_0)\}$ to the first two terms 
 of the asymptotic expansion (\ref{fitgamma}) cannot lead to  
 very accurate estimates of $\gamma(g_0)$, because, 
 even at the present order of expansion, the contribution
 of higher-order corrections is not  sufficiently small.
 In the figures \ref{fig_3_phi43}  and \ref{fig_4_phi43}, 
 we have represented by solid lines the estimates
$\gamma(g_0)$ obtained from the MRAs $\gamma^{(17)}(g_0)$,  
 $\gamma^{(21)}(g_0)$ and $\gamma^{(25)}(g_0)$, which use the 
susceptibility series only up to the orders $17$, $21$ and $25$ respectively. 
 These estimates show a rapid crossover from the Gaussian value $\gamma(0)=1$
 to a  behavior which, for $g_0 > 0$, tends to become independent of $g_0$  
 and to approach the dashed line  in the figure  indicating
  the central value $\gamma= 1.2371$ 
 of our estimate in Ref.\cite{bcesse25} 
  for the  Ising universality class. 
If we include  higher correction terms   
in the asymptotic expansion (\ref{fitgamma})
 somewhat more accurate estimates  of $\gamma(g_0)$ 
 can be obtained.
 The simplest possibility is to
 introduce a single effective higher
  correction   $O(1/n^s)$ with $g_0$-independent
 exponent $s$. Choosing the value $s=4$, 
 we obtain the estimates represented by the dotted curve.
The accuracy of the approximation 
can be further improved at the expense of the simplicity of the fitting 
procedure, but here it is sufficient to indicate only the qualitative trend. 
In the same figures, we 
 have also reported  estimates of $\gamma(g_0)$ obtained by simple  
first-order unbiased DAs using all available series coefficients.  
 In order to keep the figures readable, 
 these data are  restricted  to a smaller range of values of $g_0$ 
of particular interest.  
Completely analogous results are 
obtained starting with the  HT series for  the correlation-length squared
 $\xi^2$ to determine 
 the critical exponent $\nu$  but, for brevity, 
they  will not be reported here.

\subsection {Leading corrections to scaling and accurate estimates}

\begin{figure}
\includegraphics[width=3.1in]{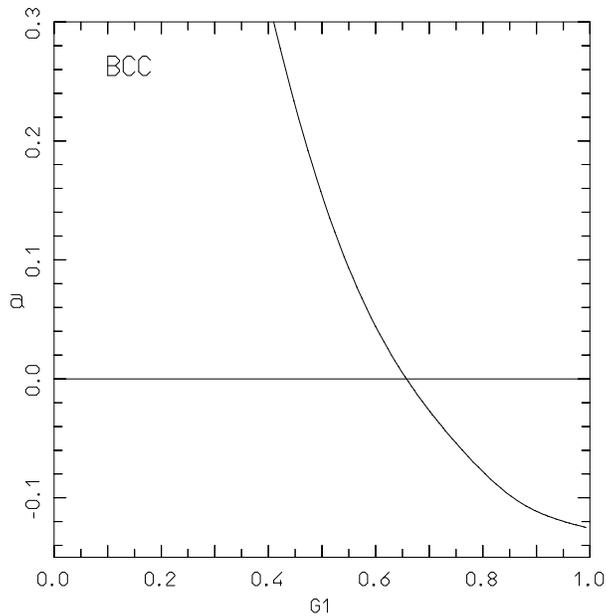}
\caption{\label{fig_5_phi43}
The amplitude $a(g_0)$  of the leading correction to scaling 
in (\ref{Suscrit}) vs. $G1= \frac {g_0} {g_0+1}$, as obtained from a
 fit to the asymptotic expansion (\ref{fitbeta})
supplemented by  a single  higher-order correction $O(1/n^{6+\theta})$ 
in the bcc lattice case.}
\end{figure}

In Fig. \ref{fig_5_phi43} we have plotted vs. $g_0$ the 
amplitude $a(g_0)$ of  the LCS in (\ref{Suscrit}), as obtained by 
 fitting  the MRA sequence $\{K^{(n)}_c(g_0)\}$  to the asymptotic expansion
 (\ref{fitbeta}) in the bcc lattice case. Of course, 
the determination of $a(g_0)$
 is much more sensitive than $K_c(g_0)$ to the inclusion of 
higher-order corrections 
in the fit to the expansion (\ref{fitbeta}). 
The most convincing results are obtained when
 a single additional correction $O(1/n^{6+\theta})$ is included. 
This three-parameter fit changes the results of the two-parameter fit 
 for $K_c(g_0)$ only within the estimated uncertainties
 and does not alter the qualitative structure of $a(g_0)$, 
 but only brings its evaluation into closer agreement with
 the corresponding estimate of $a(g_0)$
 from a similarly improved fit to eq.(\ref{fitgamma}) and with
 the values of $a(\infty)$  obtained in earlier 
 works\cite{nr90,bcesse25}. The uncertainty of the final results should not
 exceed $5-10\%$.  
In the case of the sc lattice, the  behavior of $a(g_0)$ 
 is   similar, but 
$a(g_0)$  cannot be determined with comparable accuracy 
by this straightforward  method because it is even more sensitive 
to the presence of  higher-order corrections in (\ref{fitbeta}).
The important fact is however  
 that,   both in the sc and the bcc lattice cases, 
 when $g_0$ increases from $0$ to $\infty$, 
 the amplitude $a(g_0)$ of the LCS in eq.(\ref{Suscrit})  
varies from positive to negative values.
 In the sc lattice case $a(g_0)$ vanishes at  
$\hat g^{sc}_0 \approx 1.10(2)$.  
We recall that the above cited numerical analysis of the $\phi^4_3$  HT series 
 on the sc lattice\cite{pisa25}
 was  indeed performed  precisely at $g_0 =1.10$.
  In the bcc lattice case we find that 
the zero of  $a(g_0)$ occurs at  $\hat g^{bcc}_0 \approx 1.85(5)$.
 In Ref.\cite{pisa25} $\hat g^{sc}_0$ had been 
 determined by an appropriate
 MonteCarlo simulation. 
 In our approach no preliminary MonteCarlo simulation is needed
 to determine $\hat g^{bcc}_0 $, since, knowing the structure of the function 
$a(g_0)$,   our study of the model on
 the whole range of values of $g_0$ directly gives a sufficiently
 accurate indication.
For  $g_0$ in a neighborhood of the zero of the  LCS amplitude, 
  we can observe that the various approximations of $\gamma(g_0)$, 
 shown in Figs. \ref{fig_3_phi43}  and \ref{fig_4_phi43},  
tend to coincide and come 
nearest to our earlier estimate\cite{bcesse25} for the spin $S$ Ising model: 
 $\gamma= 1.2371(1)$  shown in the figure 
 by a dashed line.
 Eq.(\ref{fitgamma}) and the measured behavior of $a(g_0)$ simply 
explain why, 
for $g_0 < \hat g^{sc}_0$, 
(or $g_0 < \hat g^{bcc}_0$ in the bcc lattice case) the MRAs  
 of Figs. \ref{fig_3_phi43} and \ref{fig_4_phi43} approach
 their constant limiting value  from below  and otherwise from above. 
Of course, the  same mode of approach to the limit can be observed
 for the estimates obtained from DAs, when varying the number of HT series
 coefficients used in the approximants, however, for clarity  
we have reported in the figures only the highest order results.

We can now take advantage of our estimate of  $\hat g^{bcc}_0 $
 to compute some universal critical parameters of the 
$\phi^4_3$ model on the bcc lattice with approximately vanishing LCS.
The  HT series coefficients of
$\chi_2$,   $\mu_2$ and $\chi_4$ on the bcc lattice,
 for $ g_0=\hat g^{bcc}_0$,
  are reported in Table II. (The corresponding expansions in the sc
 lattice case, for  $ g_0=\hat g^{sc}_0$, 
can be found in Ref.\cite{pisa25}.)
 Very accurate estimates are thus obtained for
 the critical exponents and the critical renormalized coupling.
 In particular,  using first- and second-order DAs 
(either unbiased  or biased with the critical 
 value  $K_c(\hat g^{bcc}_0)=0.2441357(5) $), 
we find $\gamma=1.2373(2) $, $\nu=0.6301(2) $, 
 $g=23.56(3) $ in good
 agreement with our previous estimates\cite{bcesse25} and with the 
results of Ref.\cite{pisa25}.
Our estimated errors  
  account  also for
 the $\approx 3 \% $ uncertainty 
  in the estimate of $ \hat g^{bcc}_0$.
 
Since our HT series for $g(K,g_0)$ 
is two terms longer, we have also repeated the analysis of Ref.\cite{pisa25}
 with our sc lattice series computed
 at $\hat g^{sc}_0=1.10$. 
The final estimates:    
$\gamma= 1.2372(2)$, $\nu= 0.6301(2)$ and  $g=23.56(4)$,
 evaluated at $K_c(\hat g^{sc}_0)= 0.375097(1)$, 
 are consistent  both with our results  for the bcc lattice and 
 with those of Ref.\cite{pisa25}.

\subsection {The renormalized coupling}

\begin{figure}
\includegraphics[width=3.2in]{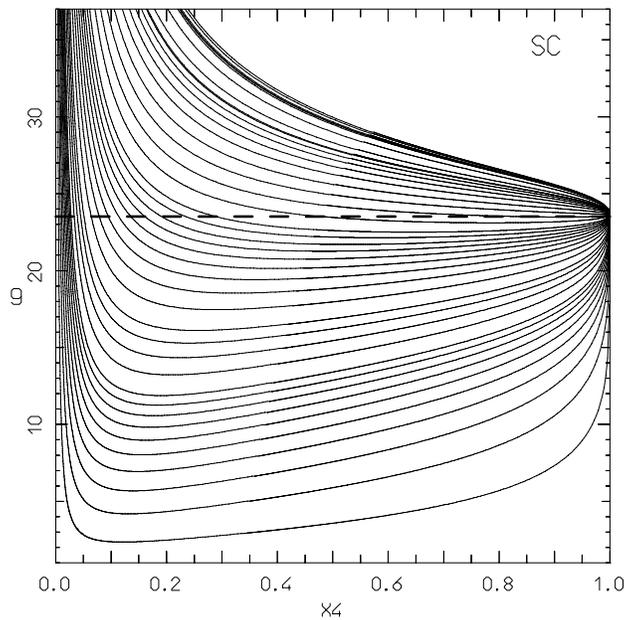}
\caption{\label{fig_6_phi43} The renormalized coupling 
  $g(\xi^2,g_0)$ as  function of $X4= \frac {\xi^2}{\xi^2+4}  $ for several 
fixed  values of the $\phi^4$ self-coupling $g_0$   
on the sc lattice. The values of $g_0$ increase from 
the lowest curve up. The dashed line represents the central value of our
 previous estimate $g=23.52(5)$, obtained in Ref.\cite{bcesse25},  
of the renormalized coupling in the Ising universality class.}
\end{figure}

\begin{figure}
\includegraphics[width=3.2in]{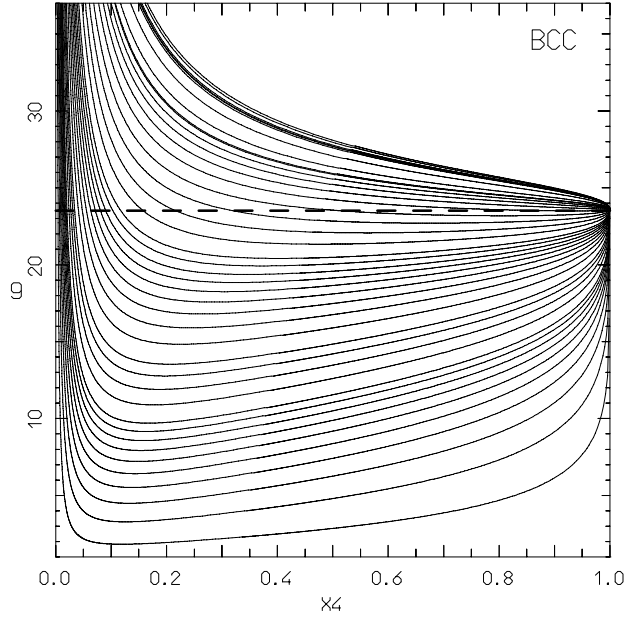}
\caption{\label{fig_7_phi43} 
The same as in Fig.\ref{fig_6_phi43}, but for the bcc
lattice.}
\end{figure}

In Figs. \ref{fig_6_phi43} 
and \ref{fig_7_phi43}, which refer to the sc and the bcc lattices 
respectively, we have plotted 
 the renormalized four-point coupling constant $g(\xi^2,g_0)$ 
 vs. $X4=\frac {\xi^2} {\xi^2+4}$ 
for several fixed values of  $g_0$.
 For fixed $g_0$ and 
 $\xi^2 \rightarrow \infty$, all curves  appear to tend to a  
$g_0$- and lattice-independent limiting value $g(\infty,g_0)$ which is 
consistent with our  estimate in Ref.\cite{bcesse25}
 of the critical renormalized
 coupling in the Ising universality class $g=23.52(5)$ indicated 
in the figure by the dashed line. 
This fact suggests that the critical 
renormalized coupling is  universal with
 respect to the self-coupling $g_0$ 
and to the structure of the lattice and moreover that it is 
non-zero (so that hyperscaling is valid).
The curves in the figures 6 and 7 are obtained in a most straightforward way 
from the highest-order non-defective ``simplified
 differential approximants''(SDA) \cite{bcsda} 
biased with the values of $\theta$
 and $K_c(g_0)$. Completely equivalent results are obtained
also, slightly more laboriously, using ordinary biased DAs instead 
of SDAs.
 The estimated width of the error bars is comparable to the
 thickness of the lines. Also for $g(\xi^2,g_0)$, we can 
observe that, as $g_0$
 varies from  $0$ to $\infty$, the amplitude  $c(g_0)$  of the LCS
  in (\ref{grencrit}), which  is negative for small  $g_0$,
 changes sign at $\hat g_0$, as it should,
 because the ratio  $c(g_0)/a(g_0)$ is expected to be universal.
Correspondingly, as shown clearly by these figures,
 the curves tend to their common limiting value 
from below if $g_0 < \hat g_0$ and otherwise from above.

\begin{figure}
\includegraphics[width=3.3in]{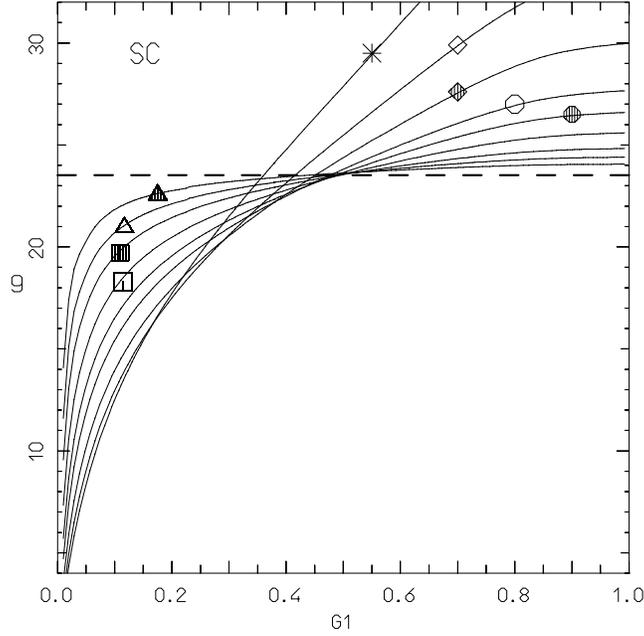}
\caption{\label{fig_8_phi43}
The renormalized coupling 
  $g(\xi^2,g_0)$ as  function of $G1= \frac {g_0}{g_0+1}  $ for several 
  values of the correlation-length square $\xi^2$   on the sc lattice. 
The  values of $\xi^2$  are indicated by the symbols
 attached to the corresponding curves.The asterisk labels
the curve for $\xi^2=1$, the empty rhomb  for $\xi^2=2$, the
full  rhomb for $\xi^2=4$, the empty circle for $\xi^2=9$,
 the full circle for $\xi^2=16$, the empty square for $\xi^2=36$, the
 full square for $\xi^2=100$, the empty triangle for $\xi^2=256$,
 the full triangle for $\xi^2=900$. 
 The dashed line represents the central value of our
 estimate $g=23.52(5)$, obtained in Ref.\cite{bcesse25}  
of the renormalized coupling in the Ising universality class.}
\end{figure}

\begin{figure}
\includegraphics[width=3.3in]{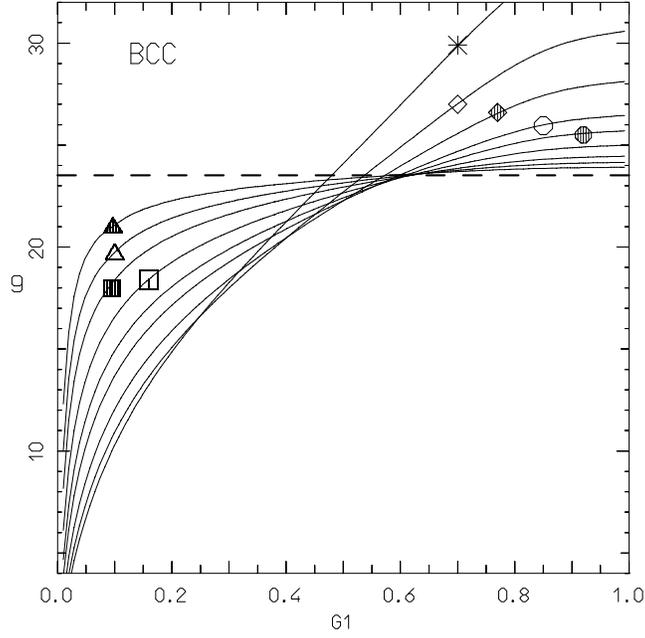}
\caption{\label{fig_9_phi43}
The same as in Fig.\ref{fig_8_phi43}, but for the bcc
lattice.}
\end{figure}

Figs. \ref{fig_8_phi43} and \ref{fig_9_phi43}, 
for the sc and the bcc lattices respectively, offer a different
view of the same data. Here the renormalized coupling 
$g(\xi^2,g_0)$  is plotted 
vs.  $G1=\frac {g_0} {g_0+1}$ 
for several fixed values of $\xi^2$.
 All curves are monotonically increasing in $g_0$  and show a rapid crossover 
from the Gaussian value $g(\xi^2,0)=0$
 to a shape increasingly flatter as $\xi^2$ becomes large.
For $g_0 < \hat g_0$, as $\xi^2 \rightarrow \infty$, 
 the sequence of curves tends from 
 below to the constant  $g(\infty,g_0)$,
  for $g_0 > \hat g_0$  the same value is approached  from above.
 We have already stressed, that this pattern of behavior simply reflects the 
fact that the amplitude $c(g_0)$  of the  LCS in $g(K,g_0)$ is negative for  
$g_0 < \hat g_0$ and positive  for  $g_0 > \hat g_0$.
 Here we should also recall that in a similar plot reported in  Ref.\cite{bin}
 (only the case of the bcc lattice case is discussed in that study)
 the curves  show  quite a different structure for $\xi^2 > 64$.
In particular, the curves $g(\xi^2,g_0)$ in  Ref.\cite{bin}
  are not monotonic in  $g_0$
 and their  limiting behavior as $\xi^2 \rightarrow \infty$
 does not appear to be universal
 and non-zero. We can  see no  reason for
 these anomalous features other than the insufficient length 
of the ten-term HT series used in the analysis of Ref.\cite{bin}.

\begin{figure}
\includegraphics[width=3.3in]{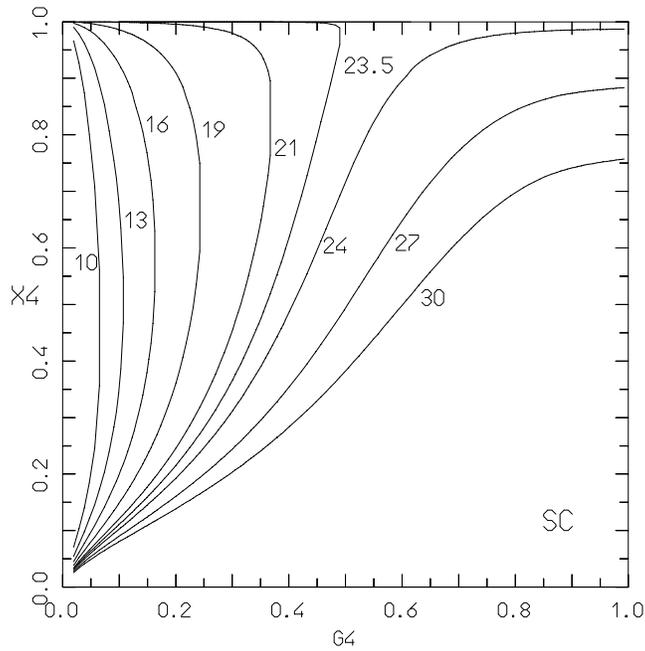}
\caption{\label{fig_10_phi43}
Contour plot  of the renormalized coupling  $g=g(\xi^2,g_0)$  
in the  $G_4$, 
$X_4$ plane  for  the sc lattice. Here $G_4= \frac {g_0}{g_0+4}  $ and
 $X_4= \frac {\xi^2}{\xi^2+4}  $. The values of $g(\xi^2,g_0)$ 
are shown beside the corresponding level curves.}
\end{figure}

\begin{figure}
\includegraphics[width=3.3in]{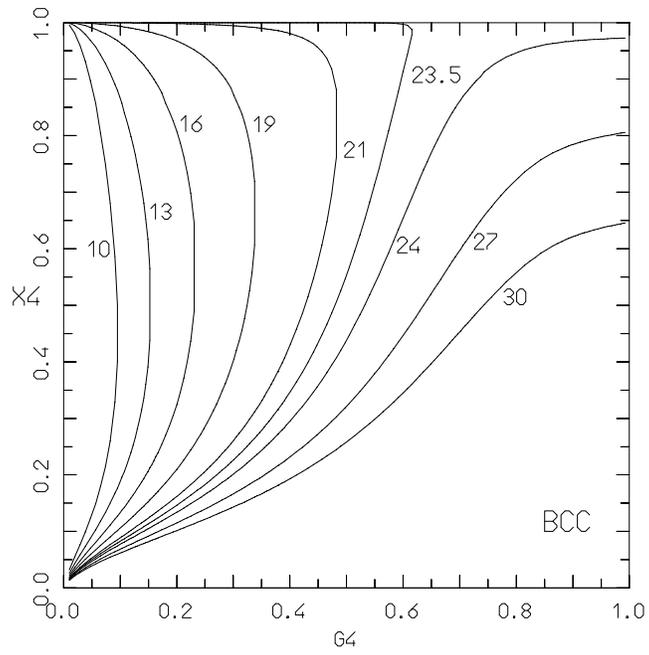}
\caption{\label{fig_11_phi43}
The same as in Fig.\ref{fig_10_phi43}, but for the bcc
lattice.}
\end{figure}

One more view of the same data is presented in Figs. 
\ref{fig_10_phi43} and \ref{fig_11_phi43},
for the sc and the bcc lattices respectively,
 showing contour plots of $g(\xi^2,g_0)$ in the $\hat \xi^2=
\frac {\xi^2} {\xi^2+1}$, $\hat g_0=
\frac {g_0} {g_0+1}$ plane. Also these figures are qualitatively different
 from the corresponding ones of Ref.\cite{bin} which 
show  a spurious  saddle-point structure 
and   suggest a failure of universality. Nothing like that can be 
inferred from our updated figures.

\subsection {Conclusions}

In conclusion,  we have analyzed HT expansions for a one-parameter 
family of continuous spin models which interpolate between the Gaussian 
and the spin-1/2 Ising models.  In the case of the  bcc lattice, we 
have taken advantage of the parameter dependence of the 
amplitudes of the LCS to improve the 
 accuracy in the determination of some universal critical quantities. 
 Moreover, we have shown that, in the light of our extended series, 
all puzzling and unexpected features which emerged from the old HT  analysis 
of Ref.\cite{bin} can only be ascribed
 to  numerical inaccuracies deriving from the use of too short series. 

\acknowledgments
The second named author (M.C.)  passed away
 before the final text of this report was completed, therefore 
the first author is entirely responsible for any errors or omissions.

This work has been partially supported by the MIUR.

\newpage

\begin{table}
\squeezetable
\caption{ $K_c(g_0)$ for a few values of
of $g_0$ in the case of the sc and bcc lattices.}
\label{tabella1}
\begin{tabular}{ccc}
\hline   
$g_0$ &  $K^{sc}_c(g_0)$   & $K^{bcc}_c(g_0)$  \\
\hline  
          0.000&       0.3333333  &        0.250000   \\
          0.010&       0.34055(1) &        0.25471(1) \\
          0.050&       0.35940(1) &        0.26724(1) \\
          0.100&       0.37340(1) &        0.27655(1) \\
          0.150&       0.38238(1) &        0.28242(1) \\
          0.200&       0.38842(1) &        0.28627(1) \\
          0.250&       0.39250(1) &        0.28876(1) \\
          0.300&       0.39517(1) &        0.29029(1) \\
          0.350&       0.396782(4) &       0.29108(1) \\
          0.400&       0.397585(3) &       0.29131(1) \\
          0.500&       0.397397(3) &       0.29055(1) \\
          0.650&       0.394135(3) &       0.28739(1) \\
          0.760&       0.390343(3) &       0.284131(4) \\
          0.900&       0.384503(3) &       0.279335(3) \\
          1.060&       0.377035(2) &       0.273375(3) \\
          1.100&       0.375096(2) &       0.271844(2) \\
          1.140&       0.373137(3) &       0.270306(2) \\
          1.300&       0.365220(3) &       0.264137(2) \\
          1.600&       0.350584(3) &       0.252883(1) \\
          1.800&       0.341319(3) &       0.245830(1) \\
          1.850&       0.339085(3) &       0.2441357(5) \\
          2.000&       0.332596(3) &       0.239230(1) \\
          3.000&       0.297990(3) &       0.213328(1) \\
          4.100&       0.274459(3) &       0.195919(1) \\
          5.000&       0.262538(3) &       0.187155(2) \\
          6.400&       0.251303(3) &       0.178926(2) \\
         10.000&       0.238853(4) &       0.169848(2) \\
         15.000&       0.232587(4) &       0.165296(2) \\
         45.000&       0.225112(4) &       0.159875(2) \\
\end{tabular} 
\end{table}

\begin{table}
\squeezetable
\caption{ The HT expansions of the susceptibility $\chi_2$, 
the second moment of the correlation-function
  $\mu_2$ and the zero-momentum four-point function $\chi_4$ 
 evaluated at $g_0=1.85$   on the bcc lattice.}
\label{tabella2}
\begin{tabular}{cccc}
\hline   
$order$ & $\chi_2$   & $\mu_2$& $\chi_4$  \\
\hline  		      
  0& 0.607989217187788034965& 0.000000000000000000000&-0.530149722377688222012\\
  1& 2.957207105732954317729& 2.957207105732954317729&-10.31441007042348207915\\  
  2& 13.09429900741101048493& 28.76720053242789148965&-116.8619663540111359343\\  
  3& 57.79323569148811046867& 197.7144175506477337902&-1037.709279795438675000\\
  4& 248.2179229506115615651& 1185.408945561549742349&-7930.113134914392838073\\
  5& 1064.858590911481926754& 6551.599918767001108148&-54882.22830020753646424\\
  6& 4512.534104619384245591& 34421.02131513359339522&-353349.3513943436529616\\
  7& 19111.25951155995320831& 174321.7172035645509269&-2155038.392864457644076\\
  8& 80383.21948173478800467& 859681.0962551252453637&-12593879.17023276266910\\
  9& 337958.2776471983168127& 4151521.352493417270411&-71122511.16749072143498\\
 10& 1414677.216907545784892& 19721977.30824040566487&-390463929.5724427479526\\
 11& 5919934.511772996630729& 92423661.55059636566134&-2093673545.070737911244\\
 12& 24698142.58226079299701& 428322758.4354179679972&-11002848880.38386524575\\
 13& 103016201.6436073223436& 1966128942.161860887810&-56833474652.32522843381\\
 14& 428735652.9640138248084& 8952466557.867528601072&-289181057575.1432464575\\
 15& 1783970290.009394230477& 40476319807.17866188775&-1452144744869.861166239\\
 16& 7410672453.259276817146& 181885433500.9131802296&-7207354502435.721114810\\
 17& 30779127150.29510373162& 812879825813.3598744239&-35401801340419.05097854\\
 18& 127668343287.7040888522& 3615487616835.276408009&-172273869711328.5865152\\
 19& 529480497190.9511942304& 16011252152185.13150869&-831302587175201.5552329\\
 20& 2193593476094.224975766& 70632026759664.65740219&-3980914502340770.007282\\
 21& 9086796325630.050694943& 310488939709344.3260940&-18931663273718966.90898\\
 22& 37608562966496.06439966& 1360524482339925.373685&-89460744874166538.46043\\
 23& 155638849919474.9034983& 5944225695917982.872040&-420283001120399961.6139\\
 24& 643624490532272.2532159& 25901506033628787.27895&\\
 25& 2661386657660504.869714& 112585550276346019.8717&\\
\end{tabular} 
\end{table}

\end{document}